\documentclass[amsmath,amssymb,superscriptaddress,prl,reprint,widetext,longbibliography]{revtex4-1}

\usepackage[utf8]{inputenc}
\usepackage{graphicx}
\usepackage{xcolor}
\usepackage{hyperref}
\usepackage{notes2bib}
\usepackage{siunitx}
\DeclareSIUnit\Molar{\textsc{m}}

\begin{document}

\title{Recoil experiments determine the eigenmodes of viscoelastic fluids}

\author{F\'elix Ginot}
\affiliation{Fachbereich Physik, Universit\"{a}t Konstanz, 78457 Konstanz, Germany}

\author{Juliana Caspers}
\affiliation{Institute for Theoretical Physics, Georg-August Universit\"{a}t G\"{o}ttingen, 37073 G\"{o}ttingen, Germany}

\author{Luis Frieder Reinalter}
\affiliation{Fachbereich Physik, Universit\"{a}t Konstanz, 78457 Konstanz, Germany}

\author{Karthika Krishna Kumar}
\affiliation{Fachbereich Physik, Universit\"{a}t Konstanz, 78457 Konstanz, Germany}

\author{Matthias Krüger}
\affiliation{Institute for Theoretical Physics, Georg-August Universit\"{a}t G\"{o}ttingen, 37073 G\"{o}ttingen, Germany}

\author{Clemens Bechinger}
\email{clemens.bechinger(at)uni-konstanz.de}
\affiliation{Fachbereich Physik, Universit\"{a}t Konstanz, 78457 Konstanz, Germany}

% ========================== Abstract ==========================
\begin{abstract}
We experimentally investigate the recoil dynamics of a colloidal probe particle after shearing it with constant velocity through a viscoelastic fluid. The recoil displays two distinct timescales which are in excellent agreement with a microscopic model built on a particle being linked to two bath particles by harmonic springs. This model yields analytical expressions which reproduce all experimental protocols, including additional waiting periods before particle release. Notably, two sets of timescales appear, corresponding to reciprocal and nonreciprocal eigenmodes of the model.  

\end{abstract}

\maketitle

% ========================= Introduction ========================

The rheological properties of complex fluids such as polymer and micellar solutions, colloidal suspensions and more, are of central importance for many natural phenomena and technical applications on macroscopic and microscopic length scales. 
Compared to purely viscous, i.e. Newtonian fluids, their response to stress or strain exhibits a non-trivial time-dependence which arises due to their mesoscopic microstructure~\cite{de1976dynamics,doi1988theory,wang2013new}. 
This allows to store and dissipate energy on rather long (up to several seconds) relaxation times~\cite{adam1983viscosity,ashwin2015microscopic}. To characterize such local relaxation processes, microrheology has emerged as a powerful technique since it allows to exert only local perturbations to the fluid using e.g. colloidal probe particles~\cite{mason1997particle,levine2000one,squires2010fluid,wilson2011small,robertson2018optical}. 
In particular when these probes are driven by steady or oscillating external fields (active microrheology), the non-linear mechanical response of complex fluids can be characterized~\cite{liu2006microrheology,chapman2014onset,neckernuss2015active,fitzpatrick2018synergistic,ge2018nanorheology,jain2021two}.

Recent experiments with micron-sized colloidal probes dragged through different types of complex fluids (wormlike micelles, polymer solutions, and entangled $\lambda$-phage DNA) with an optical trap, revealed a rich transient recoil dynamics after the trap was removed~\cite{wilking2008optically,chapman2014nonlinear,Gomez-Solano2015-qu,falzone2015entangled,zhou2018dynamically,Khan2019}. 
Notably, experiments with slightly different shearing protocols demonstrated  different relaxation behaviors, even when the same viscoelastic material was studied. 
This raises the question, what material properties are actually associated with the measured timescales and how such recoil experiments depend on the specific choice of the experimental protocol.

In this work we experimentally investigate the relaxation (recoil) of a colloidal bead after driving it through a wormlike micellar solution with an optical tweezers and then suddenly turning it off (particle release). 
Independent of the shearing protocol, the recoil always proceeds via a double-exponential relaxation process.
While the two timescales are independent of the applied protocol, the associated amplitudes strongly depend on the details of how the probe is driven through the fluid.  
Our experimental results are in good agreement with a microrheological model where the response of the fluid is described with two \emph{bath} particles connected by linear springs. 
This model has two sets of eigenmodes, corresponding to reciprocal (trap off) and nonreciprocal forces (trap on), whose excitation depends on the protocol. 
These findings may explain the above mentioned discrepancies seen in previous studies.

% =========================  Experimental Setup ========================

In our experiments we used a viscoelastic solution of \SI{5}{\milli\Molar} equimolar cetylpyridinium chloride monohydrate (CPyCl) and sodium salicylate (NaSal) to which we added a small amount of silica probe particles with diameter $\SI{2.73}{\micro\m}$.
The solution was contained in a sealed sample cell with \SI{100}{\micro\m} height being kept at a temperature of \SI{25}{\celsius}.
Under such conditions, the fluid forms an entangled network of giant worm-like micelles with pronounced viscoelastic properties~\cite{cates1990statics}. The colloidal probe was optically trapped in the focus of a Gaussian laser beam $\lambda=\SI{1064}{\nano \m}$ using an high magnification microscope objective (100x, NA= 1.45).
This yielded a rather stiff trap with trapping strength $\kappa_{\rm OT}=32\pm \SI{1}{\micro\N/\m}$.
To avoid possible interactions with the sample walls, the trap was located at least \SI{30}{\micro \m} away from any surface.
Motion of the probe relative to the fluid has been achieved, by translating the sample cell with constant velocity using a computer controlled piezo-driven stage.
This motion has been synchronized with the laser intensity to realize different shearing protocols as described in detail below.
Pictures of the probe particles have been recorded by video microscopy with a frame rate of \SI{100}{\hertz}.
Using a custom Matlab \cite{crocker1996methods} algorithm, the particle position has been resolved with an accuracy of $\pm \SI{6}{\nano\m}$.
For further details regarding the experimental setup, we refer to the Supplementary Material (SM)~\cite{SMref}.

% =========================  Figure 1 - Sketch of Experiment ========================

\begin{figure}
    \centering
    \includegraphics{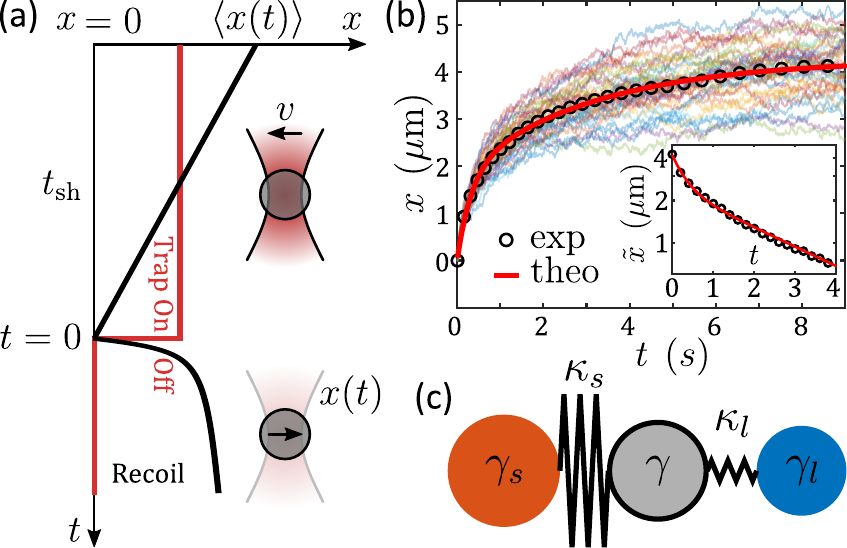}
    \caption{(a) Sketch of a typical recoil protocol. The colloidal probe is first driven by an optical tweezers through the fluid along the x-axis for a time $t_\text{sh}$ and at a constant velocity $v$. At $t=\SI{0}{\s}$ the optical trap is turned off and the probe is released. Because of accumulated strain in the fluid, the particle performs a recoil opposite to the direction of shear. (b) Typical recoil curves (colored lines) obtained for $t_\text{sh}=\SI{8}{\s}$ and $v =\SI{3}{\micro\m\per\s}$ and the mean value $\langle x(t)\rangle $ (black circles). The red solid line is a double-exponential according to Eq.~\eqref{eq:BiExpFit} (red line). Inset: $\Tilde{x}(t)=A_\text{tot}-\langle x(t)\rangle$ in lin-log scale highlights the two timescales. (c) Sketch of the two-bath particle model where the probe with friction coefficient $\gamma$ is linearly coupled to two bath particles.
    }
    
    \label{fig:F1}
\end{figure}

Fig.~\ref{fig:F1}a shows a schematic of the experimental protocol used in our study.
A probe particle is first trapped by an optical tweezers and dragged with constant velocity $v$ through the solution for a time $t_\mathrm{sh}$. 
At time $t=\SI{0}{\s}$, the optical trap is turned off and the particle experiences a recoil opposite to the direction of $v$ along the x-axis. 
In Fig.~\ref{fig:F1}b we plotted the results of such an experiment with $v= \SI{3}{\micro\m\per\s}$. 
The shear time was set to $t_\mathrm{sh}=\SI{8}{\s}$ which is sufficiently long that the system approaches a non-equilibrium steady state where the recoils become independent of $t_\mathrm{sh}$. 
Note that due to particle sedimentation the particles eventually disappear of the imaging focal plane which limits individual recoil experiments to about $\sim \SI{10}{\s}$. 
Because of thermal noise, the individual trajectories (colored thin lines) scatter around the corresponding mean value which has been obtained from about 100 repetitions of the protocol (black open symbols).
As shown in Fig.~\ref{fig:F1}b and in agreement with previous studies, the mean recoil $\langle x(t) \rangle $ is well described by a superposition of two exponential (red plain line) relaxation processes~\cite{Gomez-Solano2015-qu}
\begin{equation}
	\langle x(t)\rangle = A_\mathrm{tot} -   A_s e^{-\frac{t}{\tau_{s}}} - A_l e^{-\frac{t}{ \tau_{l}}}, \label{eq:BiExpFit}
\end{equation}
where $\tau_\text{s}$ and $\tau_\text{l}$ are two timescales with amplitudes $A_s$ and $A_l$ respectively, and where $A_\mathrm{tot}=A_s+A_l$ is the total recoil amplitude.
The presence of two timescales is better highlighted in the inset of Fig.~\ref{fig:F1}b, where we plotted $\Tilde{x}(t)=A_\text{tot}-\langle x(t)\rangle$ in a lin-log scale.

% =========================  Figure 2 - Vary V  ========================

\begin{figure}
    \centering
    \includegraphics{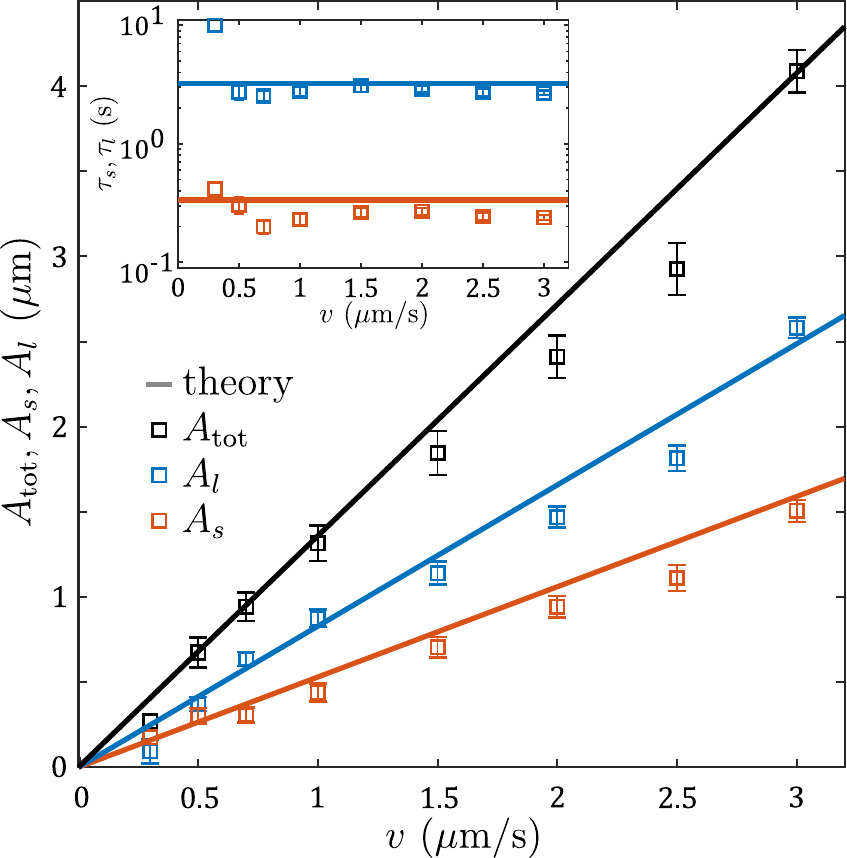}
    \caption{Recoil amplitudes $A_\text{tot}$ (black), $A_s$ (red), and $A_l$ (blue) as a function of the shear velocity $v$, for $t_\mathrm{sh}=\SI{8}{\s}$.
    The experimental data (squares) are obtained by fitting the individual recoil curves (see Fig.~\ref{fig:F1}b) with Eq.~\eqref{eq:BiExpFit}. Theoretical curves (lines) are calculated from Eq.~\eqref{eq:AslAnalytic}.
    Inset: Relaxation times $\tau_{s}$ (red) and $\tau_{l}$ (blue) extracted from the recoil curves. Square symbols correspond to experiments, and lines to theoretical predictions according to Eq.~\eqref{tauSLAnalytic}.
    }

    \label{fig:F2}
\end{figure}

To phenomenologically describe the observed recoil behavior and in particular the occurrence of two timescales in the relaxation process we consider a simple model which is shown in Fig.~\ref{fig:F1}c. 
Here, the probe particle with friction coefficient $\gamma$ is coupled via linear springs with stiffness $\kappa_s$ and $\kappa_l$, respectively, to two bath particles with friction coefficients $\gamma_s$ and $\gamma_l$. 
Such model corresponds to an extension of the well known Maxwell model where a single harmonically coupled bath particle is considered, and which is known to provide a good description for the equilibrium properties of viscoelastic fluids \cite{Rehage1988,cates1990statics,Walker2001,Dreiss2007,doerries2021correlation}. 

To rationalize the assumption of a harmonic coupling of the bath particles to the probe, we investigated how $A_\mathrm{tot}$ varies with  $v$. 
Within our model, a variation of the shear velocity leads to changes in the elastic forces between the probe and the bath particles, therefore such experiments allow to investigate the properties of the considered springs. 
From our experiments we observe that all recoil amplitudes $A_s$, $A_l$, $A_\mathrm{tot}$ are proportional to $v$ (Fig.~\ref{fig:F2}). 
Notably, the corresponding relaxation times are independent of $v$ (inset of Fig.~\ref{fig:F2}). 
As discussed further below, these observations suggest the choice of a linear model, in particular the assumption of linear springs in accordance to Fig.~\ref{fig:F1}c.

Clearly, the validity of our model is limited to small shear velocities since it does not describe the saturation of $A_\text{tot}$ at large $v$, which has been observed in previous studies where shear velocities up to $v= \SI{30}{\micro \m/\s}$ were applied~\cite{Gomez-Solano2015-qu}. 
Such saturation results from the finite amount of elastic energy which can be stored in the solvent, an effect which cannot be captured using linear springs.

The corresponding linear Langevin equations describing the dynamics of the probe and the two bath particles according to Fig.~\ref{fig:F1}c, are given by ($i\in\{s,l\}$)
\begin{align}
\begin{split}
	\gamma\dot{x}(t) &= -\kappa(t) \left[x(t)-\int^t dt'v(t')\right]  \\
	&\quad-\sum_i\kappa_i \left[x(t) - x_i(t)\right]  + \xi(t) 
	\end{split} \label{LangevinEquationTracer}\\
	\gamma_i \dot{x}_i(t) &= - \kappa_i \left[x_i(t)-x(t) \right]+\xi_i(t). \label{LangevinEquationBath}
\end{align}
The first term on the r.h.s. of Eq.~\eqref{LangevinEquationTracer} describes the interaction of the probe with the harmonic optical trap at position $x_0(t)=\int^t dt' v(t')$ during $t_\mathrm{sh}$. 
The time dependent laser trap stiffness $\kappa(t)$ equals $\kappa_{\rm OT}$ when the trap is on (during $t_\mathrm{sh}$) and zero when the trap is off.
$\xi$, $\xi_s$, and $\xi_l$ are independent Gaussian white noises, i.e., for ($(\xi_i,\xi_j)\in \{\xi,\xi_s,\xi_l\}$), 
\begin{equation}
	\left\langle \xi_{i}(t) \right\rangle = 0 \quad\text{, }\quad \left\langle \xi_{i}(t)\xi_{j}(t') \right\rangle =\delta_{ij} 2k_{\rm B}T\gamma_{i}\delta(t-t')  .  \label{NoiseCorrelation}
\end{equation}
As a first encouragement, we note that the set of Eqs.~\eqref{LangevinEquationTracer} and \eqref{LangevinEquationBath} reproduces the experimentally observed dynamics of the probe during the recoil shown in Eq.~\eqref{eq:BiExpFit} (for details see SM). 
The two timescales take the forms
\begin{eqnarray}
	\tau_{s,l}^{-1} &=&\frac{1}{2 \gamma }\left[\sum_i\zeta_i \pm 
\sqrt{(\sum_i\zeta_i)^2+4(\kappa_s\kappa_l-\zeta_s\zeta_l)}\right],\label{tauSLAnalytic}
\end{eqnarray}
where $\zeta_i=(\gamma+\gamma_i)\kappa_i/\gamma_i$.
The positive and negative signs correspond to the shorter $\tau_s$ and longer $\tau_l$ timescales, respectively. Because the two bath particles are mechanically coupled across the probe particle, these timescales depend on the combination of both stiffnesses $\kappa_{s,l}$.

Comparing Eq.~\eqref{eq:BiExpFit} with the experimental data yields the parameters 
 $\gamma_s/\gamma = 1.1$, $\gamma_l/\gamma = 0.88$, $\kappa_s/\gamma= \SI{1.5}{\s^{-1}}$, and $\kappa_l/\gamma= \SI{0.2}{\s^{-1}}$ leading to well separated relaxation times  ${\tau}_s  = \SI{0.34}{\s}$ and ${\tau}_l=\SI{3.2}{\s} $, respectively. Note, that all parameters scale with $\gamma$ which can be obtained from an independent flow curve experiment (SM).
 
\begin{figure}
    \centering
    \includegraphics{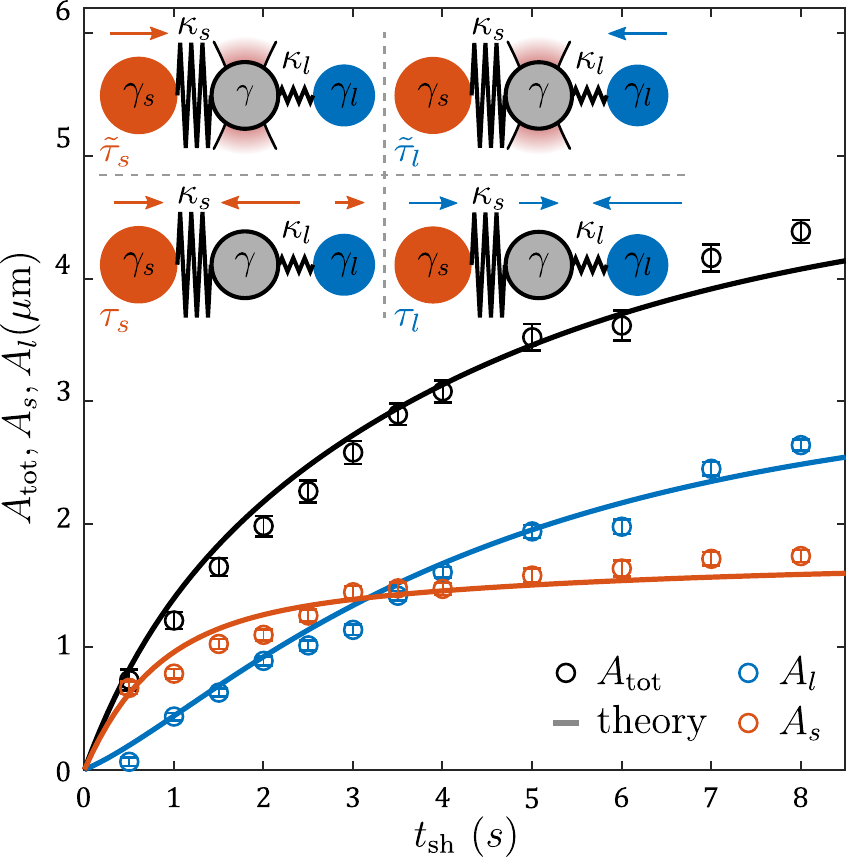}
    \caption{Recoil amplitudes as a function of the shear time $t_\mathrm{sh}$, for $v = \SI{3}{\micro\m\per\s}$ for experiments (circles) and theory (lines). For short shearing (small values of $t_\mathrm{sh}$), the short timescale dominates the recoil dynamics, i.e., $A_{s} > A_{l}$. This behavior reverses for long times where  $A_s < A_{l}$, with a cross-over at $t_\mathrm{sh} \sim \SI{3}{\s}$. Inset: Comparison between the eigenmodes associated with $\Tilde{\tau}_{s,l}$ (upper panel) when the probe is trapped at a fixed position and with $\tau_{s,l}$ (lower panel), when the probe is free. Since the shearing process is nonreciprocal, there is no coupling between the two bath particles, thus the dynamics for the short and long timescales follow $\Tilde{\tau}_{s,l}$. 
    }
    
    \label{fig:F3}
\end{figure}

To investigate, how the recoil depends on the shear time, we have repeated the above experiments for constant $v = \SI{3}{\micro\m\per\s}$ and with $t_\mathrm{sh}$ being systematically increased. 
The corresponding recoil amplitudes are shown as a function of $t_\mathrm{sh}$ in Fig.~\ref{fig:F3} (circles). 
Opposed to the dynamics of the recoil itself which proceeds again via $\tau_{s,l}$ (see SM), the characteristic timescales of the saturation behavior of the amplitudes vs. $t_\mathrm{sh}$ are given by the relaxation times of the two uncoupled bath particles, i.e. for a spatially \emph{fixed} probe particle, i.e., $\tilde\tau_s=\SI{0.7}{ \s}$ and $\tilde\tau_l=\SI{4.4}{\s}$. 
Such behavior is also in excellent agreement with our model (solid line).

These different timescales can be understood by analyzing the characteristic eigenmodes of the microscopic model (SM). 
During $t_\mathrm{sh}$ the probe's position is only determined by the optical tweezers moving at fixed velocity $v$ and not affected by the bath particles. 
As a result, the interaction between the probe and the bath particles is nonreciprocal: the two bath particles are then fully decoupled leading to their individual relaxation times $\Tilde{\tau}_{s,l}$ (upper panel in Fig.~\ref{fig:F3}). 
Once the trap is turned off, reciprocity and thus force equilibrium between all three particles must apply, resulting in more complex eigenmodes characterized by the timescales ${\tau}_{s,l}$ (lower panel in Fig.~\ref{fig:F3}). 
This is also seen in the corresponding theoretical expressions for the recoil amplitudes which depend on $\tilde\tau_i$ and $\tau_i$ ($i\in \{s,l\}$)
\begin{align}
&\frac{A_{s}}{v} = \frac{\gamma_s\tilde\tau_s\left(1-e^{-\frac{t_\mathrm{sh}}{\tilde\tau_s}}\right)}{2 \left(\gamma+\gamma_s+\gamma_l\right)}\left[1+ \frac{\tilde\tau_l \left[\zeta_l (\tilde\tau_l-\tilde\tau_s)+\gamma_s+\gamma_l \right]}{(\gamma+\gamma_s+\gamma_l)(\tau_l-\tau_s)} \right] \label{eq:AslAnalytic} \notag\\
&+\frac{\gamma_l\tilde\tau_l\left(1-e^{-\frac{t_\mathrm{sh}}{\tilde\tau_l}}\right)}{2 \left(\gamma+\gamma_s+\gamma_l\right)}\left[1- \frac{\tilde\tau_s \left[\zeta_s (\tilde\tau_s-\tilde\tau_l)+\gamma_s+\gamma_l \right]}{(\gamma+\gamma_s+\gamma_l)(\tau_s-\tau_l)} \right]. 
\end{align}
$A_l$ follows from $A_s$ by changing indices $s\leftrightarrow l$. Eq.~\eqref{eq:AslAnalytic} also confirms the experimentally observed $t_\mathrm{sh}$-dependence of the amplitudes on $\tilde\tau_{s,l}$ (see factors in large round brackets) as shown in Fig.~\ref{fig:F3}.
Notably, $A_l$ is smaller than $A_s$ at small shear time because $A_l$ contains a negative contribution, causing the curve to be rather flat at short time and inflected at later shear times. 
As a result, in this regime the recoil is largely dominated by just one timescale $\tau_s$.

\begin{figure}
    \centering
    \includegraphics{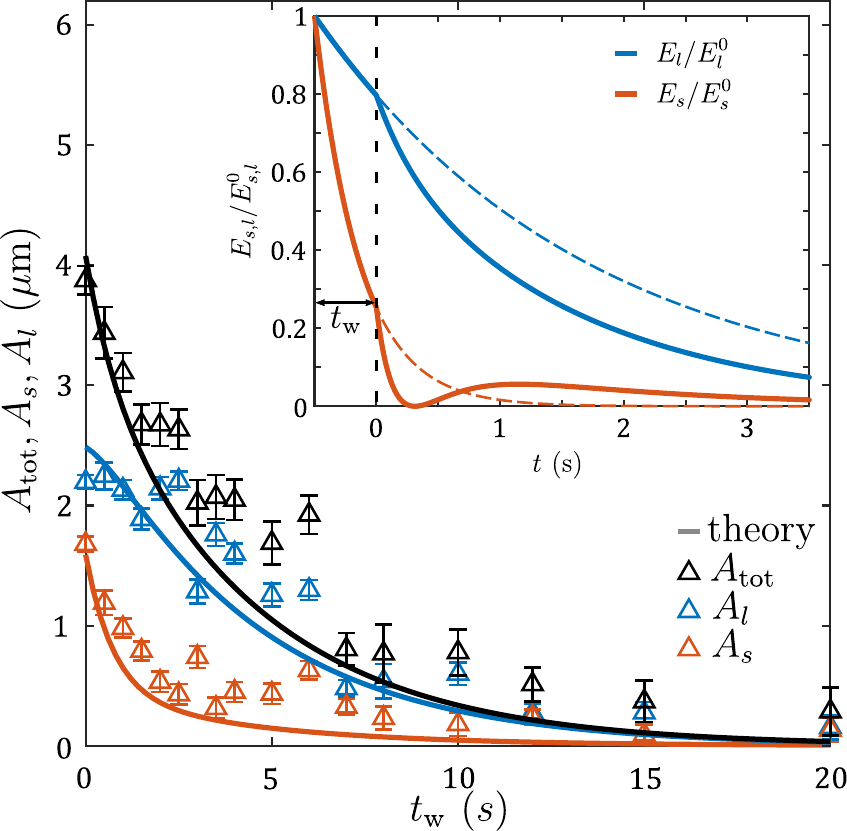}
    \caption{Recoil amplitudes as a function of the waiting time $t_w$ and initial conditions $v = \SI{3}{\micro\m\per\s}$ and $t_\mathrm{sh} = \SI{8}{\s}$. Experiments and theory are shown as symbols and lines, respectively. $A_l$ always prevails in the recoil behavior, and becomes more dominant when increasing $t_w$. Inset: Computed (normalized) energy $E_{s,l}/E_{s,l}^0$ values associated with the stiff ($\kappa_s$, red lines) and weak ($\kappa_l$, blue lines) springs during an experiment with $t_w=\SI{0.5}{\s}$. During the time $t_w$ the probe is kept at fixed position, and the interactions with the bath particles are nonreciprocal: the relaxation modes thus follow $\tilde\tau_{s,l}$. At $t=\SI{0}{\s}$ the probe is released in a force-free environment, and the system thus relaxes according to $\tau_{s,l}$. All curves were normalized by $E_{s,l}^0$, the energy associated with each spring prior to the release. Thin dashed lines show a full $\tilde\tau_{s,l}$ relaxation process, to better highlight the differences with $\tau_{s,l}$.
    }
    \label{fig:F4}
\end{figure}

%***************************************

For a direct experimental demonstration of the different relaxation behaviors depending on whether the probe is confined to the trap or not, we have changed our protocol: instead of turning off the optical trap immediately after $t_\mathrm{sh}=\SI{8}{\s}$ ($v = \SI{3}{\micro\m\per\s}$), it remained on for an additional waiting time $t_w$ but with $v=0$ (see SM for a sketch of the full protocol).
This allows the bath particles to relax independently towards the fixed probe particle, prior to the recoil where the confined probe provides a coupling between both bath particles. 
Thus, this process is expected to be fully symmetric to the above loading experiments (Fig.~\ref{fig:F3}) with the decoupled timescales of the bath particles $\tilde\tau_i$ ($i\in\{s,l\}$) (SM). 
Indeed, the measured (triangles) $t_w$-dependent amplitudes of the recoil are in good agreement with the theoretical prediction (solid lines) as shown in Fig.~\ref{fig:F4}.

Because the motion of the bath particles is accessible via Eqs.~\eqref{LangevinEquationTracer},\eqref{LangevinEquationBath}, we can also compute the elastic energies stored in the two springs ($E_l, E_s$). 
The corresponding values are shown in the inset of Fig.~\ref{fig:F4} for the above protocol during $t_w$ and for the subsequent recoil. 
As expected, the decay of $E_l$ and $E_s$ is very different during $t_w$ compared to the recoil. 
Opposed to $E_l$ which decreases monotonically, $E_s$ shows a non-monotonic behavior with a minimum around \SI{0.3}{\s}. 
This is a result of the coupling of the two bath particles (via the probe) during the recoil which leads to a partial exchange of elastic energies of the two springs.

As a final remark, we want to mention that recoil experiments provide advantages compared to equilibrium studies when analyzing the properties of viscoelastic materials. 
As a result of the sudden release of accumulated stress acting on the colloidal particle, its motion is less affected by thermal noise which leads to better signal/noise ratios compared to situations where the probe's motion is only determined by thermal equilibrium fluctuations. 
Therefor, recoil experiments allow to resolve features otherwise easily obscured by noise. 
As an example we have measured the mean-square displacement (MSD) of the probe in our micellar system (SM). 
Even though the data are well explained by our two-bath particle model, similar agreement is achieved when only a single-bath particle model is considered (corresponding to a Maxwell model)~\cite{mason_optical_1995,mason_diffusing-wave-spectroscopy_1997,van_zanten_brownian_2000,lu_probe_2002,vandergucht2003}. 
When fitting such a model (which is characterized by a single timescale) to the MSD, only the short timescale $\tau_s$ is recovered.  

In summary, we have analyzed the recoil dynamics of a colloidal particle after it was dragged through a viscoelastic fluid. 
The experimentally observed double-exponential recoil dynamics is in excellent agreement with the two eigenmodes of a linear two-bath particle model which reproduces the observed dependence of the recoil on the shear velocity, shear time and the waiting time prior to recoil. 
Depending on the specific protocol, the magnitude of slow relaxation processes is largely suppressed which may explain why single- and double-exponential recoils have been previously observed in different experiments. 

% =========================  Acknowledgments ========================
\begin{acknowledgments}
We thank Matthias Fuchs for fruitful discussions.
This project was funded by the Deutsche Forschungsgemeinschaft (DFG), Grant No. SFB 1432 - Project ID 425217212. F.G. acknowledges support by the Alexander von Humboldt foundation.
\end{acknowledgments}

% =========================  Bibliography ========================

%\bibliographystyle{apsrev4-1} 
%\bibliography{Recoil.bib} 


\end{document}